\documentclass[pra,twocolumn,superscriptaddress,showpacs]{revtex4}
\usepackage{graphicx}
\usepackage{subfigure}
\usepackage{amsmath, amsfonts, amssymb}
\newcommand{\ket}[1]{\vert #1 \rangle} \newcommand{\bra}[1]{\langle #1 \vert}

\newcommand{\ketbra}[2]{\vert #1 \rangle \! \langle #2 \vert}

\newcommand{\be}{\begin{equation}}
\newcommand{\ee}{\end{equation}}
\newcommand{\bae}{\begin{eqnarray}} \newcommand{\eae}{\end{eqnarray}}

\def\Tr{\hbox{Tr}}
\begin{document}
\title{Non-Gaussian quantum discord for Gaussian states}
\author{Paolo Giorda}
\affiliation{Institute for Scientific Interchange Foundation (ISI), I-10126 Torino, Italy}
\author{Michele Allegra}
\affiliation{Institute for Scientific Interchange Foundation (ISI), I-10126 Torino, Italy}
\affiliation{Dipartimento di Fisica, Universit\`a di Torino, I-10125 Torino, Italy}
\affiliation{INFN, Sezione di Torino, I-10125 Torino, Italy}
\author{Matteo G. A. Paris}
\affiliation{Dipartimento di Fisica, Universit\`a degli Studi di Milano, I-20133 Milano, Italy}
\affiliation{CNISM, Unit\`a di Milano, I-20133 Milano, Italy}
\begin{abstract} In recent years the paradigm based on entanglement as
the unique measure of quantum correlations has been challenged by the
rise of new correlation concepts, such as quantum discord, able to
reveal quantum correlations that are present in separable states.
It is in general difficult to compute quantum discord, because it
involves a minimization over all possible local measurements in a
bipartition. In the realm of continuous variable (CV) systems, a
Gaussian version of quantum discord has been put forward upon restricting to
Gaussian measurements. It is natural to ask whether non-Gaussian
measurements can lead to a stronger minimization than Gaussian ones.
Here we focus on two relevant classes of two-mode Gaussian states: 
squeezed thermal states (STS) and mixed thermal states (MTS), and allow 
for a range of experimentally feasible non-Gaussian measurements, 
comparing the results with the case of Gaussian measurements. We provide 
evidence that Gaussian measurements are optimal for Gaussian states.  
\end{abstract}
\maketitle
\section{introduction}  
\label{Sec : Introduction}
In recent years the paradigm based on entanglement~\cite{Horodecki} as
the unique genuine measure of quantum correlations has been challenged
by the argument that the notion of nonseparability may be insufficient
to encompass all correlations that can be fairly regarded as quantum, or
nonclassical. This has given spur to the development of conceptually new
correlation measures, such as quantum
discord~\cite{Ollivier,Vedral,DiscordRev}, based on local measurements
and able to reveal quantum correlations that are present even in
separable states.  These correlations can be interpreted as an extra
amount of information that only coherent operations can
unlock~\cite{Gu}. In fact, 
there are several indications suggesting that general quantum
correlations might be exploited in
quantum protocols~\cite{Datta2}, including mixed state quantum
computation~\cite{Datta} and remote state preparation~\cite{RSP}.
Therefore, a more complete theoretical and experimental investigation
thereof is now a central issue in quantum science and technology
\cite{Fer12}. 
\par
The definition of discord involves an optimization over all possible 
local measurements in a bipartion, the optimal measurement leading to 
a minimal value of quantum discord. 
To perform the optimization is remarkably difficult, which hampers
analytical progress in the area. This fact has led to the definition of
other correlation measures which are conceptually similar but easier to
compute, such as the geometric discord~\cite{Dakic}. In the realm of
finite-dimensional systems, where the concept of discord was first
introduced, analytic results for quantum (geometric) discord have been
obtained for pairs of qubits when the global state is in X form (in
arbitrary form)~\cite{AnalyticDiscord,Dakic}. \par 
In the realm of
continuous variable (CV) systems, initial research efforts on quantum
discord have focused on Gaussian measurements. The Gaussian quantum
discord, proposed in~\cite{GiordaGaussDiscord,AdessoDatta}, is defined
by restricting the minimization involved in the definition of discord to
the set of Gaussian POVMs~\cite{GauuPOVM} and it can be analytically
computed for Gaussian states. Its behavior in noisy channels has been
studied in Ref.~\cite{Vasile}  - where it was shown that it is more
robust than entanglement to the decorrelating effect of independent
baths and more likely to yield non-zero asympotic values in the case of
a common bath - while its relation to the synchronization properties of
detuned, correlated oscillators has been analysed in
Ref.~\cite{Zambrini}.
\par
It is natural to investigate CV quantum discord beyond Gaussian
measurements: non-Gaussian ones may indeed allow for a stronger
minimization of discord, and in this case the Gaussian discord would be
an overestimation of the true discord. Here we focus on Gaussian states
and ask whether Gaussian measurements are optimal in this case, i.e.,
\textit{whether the Gaussian discord is the true discord for Gaussian
states}. This question is relevant for two main reasons: On one hand, if
discord is a truly useful resource for quantum information
protocols~\cite{Gu,Datta2}, then it is crucial to have a reliable
estimate of its actual value. On the other hand, from a fundamental
point of view it is important to establish how different kinds of
measurements can affect correlations in quantum states. A further
motivation comes from the fact that indeed for some non-Gaussian 
states e.g., CV Werner states, non-Gaussian measurements such as photon
counting has been proven to lead to a better minimization~\cite{NonGausDisc}.
\par
The optimality of Gaussian measurements has already been proven
analytically for two-mode Gaussian states having one vacuum normal
mode~\cite{AdessoDatta}, by use of the so-called Koashi-Winter
relation~\cite{Koashi}, but no analytic argument is available in the
general case. We address the question numerically, for the case of
two-modes, upon considering two large classes of Gaussian states, the
squeezed thermal states (STS) and the mixed thermal states (MTS), and
allowing for a range of experimentally feasible non-Gaussian
measurements based on orthogonal bases: the number basis, the squeezed
number basis, the displaced number basis.  As a result, we provide
evidence that Gaussian quantum discord is indeed optimal for the
states under study.  In addition, we also investigate the CV geometric
discord~\cite{GaussianGeom}, comparing the case of Gaussian and
non-Gaussian measurements. 
\par
This work is structured as follows. In
sec. \ref{Sec : correlations} we review quantum discord and the Gaussian
version of it; in sec. \ref{Sec : NonGaussian correlations} we
thoroughly describe the basic question we want to address in this work
and introduce non-Gaussian measurements and non-Gaussian discord; in
sec. \ref{Sec : number basis}, \ref{Sec : squeezed number basis},
\ref{Sec : displaced number basis}, we present our key results
concerning non-Gaussian discord upon measurements in the number basis,
squeezed number basis and displaced number basis; in sec.\ref{Sec :
Geometric} we discuss the behavior of non-Gaussian geometric discord;
finally, sec. \ref{Sec. : conclusions} closes the paper discussing our
main conclusions.
\section{Quantum discord and Gaussian discord} \label{Sec : correlations}
Starting from the seminal works by Ollivier and Zurek~\cite{Ollivier} and
Henderson and Vedral~\cite{Vedral}, various measures of quantum
correlations which go beyond the traditional entanglement picture have
been defined~\cite{DiscordRev}.  The most common measure of such
correlations is the \emph{quantum discord}~\cite{Ollivier,Vedral}.
Let us consider a bipartite system composed of subsystems $A$ and $B$. The
total correlations in the global state are measured by the mutual
information $I(A:B) = S(\varrho_A)  + S(\varrho_B) - S(\varrho_{AB})$.
Whenever $I(A:B) > 0$, the subsystems are correlated and we can gain
some information about $A$ by measurements on $B$ only.  However, 
there is no
unique way of locally probing the state of $B$: to do it, we can perform
different local measurements, or POVMs. Any such local POVM  $\Pi_B $ is
specified by a set of positive operators  $\{ \Pi_B^x  =  M_B^x M_B^{x \dag} \}
$ on subsystem $B$ summing up to the identity, $\sum_x \Pi_B^x =
\mathbb{I} $.  When measurement result $x$ is obtained, the state of $A$
is projected onto $\varrho_A^x= \mbox{Tr}_B [ M_B^x \varrho_{AB} M_B^{x
\dag}] $.  The uncertainty on the state of $A$ before the measurement on
$B$ is given by $S(\varrho_A)$, while the average uncertainty on the
state of $A$ after the measurement is given by the average conditional
entropy $ S^\Pi (A|B) = \sum_x p_x S(\varrho_A^x)$. Their difference
\begin{equation*}
  S(\varrho_A) - S^\Pi (A|B) = S(\varrho_A) - \sum_x p_x S(\varrho_A^x)
\end{equation*}
represents the average gain of information about the state of $A$
acquired through a local measurement on $B$. The maximal gain of
information that can be obtained with a POVM, 
\begin{eqnarray}
C(A:B) = \max_{\{ \Pi \in POVM \}} [S(\varrho_A) -  S^{\Pi}(A|B) ]  = 
\label{Eq.: classicalcorr} \\ \nonumber
=S(\varrho_A) - \mbox{min}_{\{\Pi \in POVM \}} [ S^{\Pi}(A|B) ]
\end{eqnarray}
coincides with the measure of \textit{classical correlations} originally
derived in~\cite{Vedral} under some basic and natural requirements for
such a measure. Quantum discord is then defined as the difference
between the mutual information and the classical correlations:
\begin{equation}
D(A:B) = I(A:B) - C(A:B)
\end{equation}
and measures the part of the total correlations that cannot be exploited
to gain information on $A$ by a local measurement on $B$, i.e., measures
the additional quantum correlations beyond the classical ones.
\par  
It can be verified (see e.g.~\cite{Dakic}) that the classical correlations coincide with the mutual information in the system after the measurement, maximized over all possible POVMs:
\begin{eqnarray}
C(A:B)= \mbox{max}_{\{\Pi \in POVM \}} I^\Pi(A:B)
\end{eqnarray}
where $I^\Pi(A:B) = S(\varrho_A^\Pi)  + S(\varrho_B^\Pi) -
S(\varrho_{AB}^\Pi)$ and the unconditional post-measurement states are
given by $\varrho_{AB}^\Pi = \sum_x M_B^x \varrho_{AB} M_B^{x \dag} $,
$\varrho_{A}^\Pi = \mbox{Tr}_B [\sum_x M_B^x \varrho_{AB} M_B^{x \dag}]
$, $\varrho_{B}^\Pi = \mbox{Tr}_A [\sum_x M_B^x \varrho_{AB} M_B^{x
\dag}] $.  Therefore, the quantum discord coincides with the difference
between the mutual information before and after the measurement,
minimized over all possible POVMs: 
\begin{eqnarray}
 D(A:B)= \ \mbox{min}_{\{\Pi \in POVM \}} [I(A:B) - I^\Pi(A:B)]
 \end{eqnarray}
From the prevoius considerations, it is clear that  $D(A:B)=0$ if and
only if there is a local measurement $\Pi_B$ which leaves the global
state of the system unaffected: $\exists \Pi, \  \varrho_{AB} =
\varrho_{AB}^\Pi$. Such states are called \textit{quantum-classical
states} and are in the form \begin{equation} \chi_{AB} = \sum_i p_i
\varrho_{A,i} \otimes \ket{i} \bra{i} \end{equation} where $p_i$ is a
probability distribution and $ \{ \ket{i} \}$ is a basis for the Hilbert
space of subsystem $B$. For such states, there exists at least one local
measurement that leaves the state invariant and we have $I(A:B) =
C(A:B)$, which means that we can obtain maximal information about
subsystem $A$ by a local measurement on $B$ without altering the
correlations with the rest of the system.  
\par
In the realm of continuous-variable systems, the \emph{Gaussian
discord}~\cite{GiordaGaussDiscord, AdessoDatta}  is defined by
restricting the set of possible measurements in Eq.~(\ref{Eq.:
classicalcorr}) to the set of Gaussian POVMs~\cite{GauuPOVM}, and
minimizing only over this set.  The Gaussian discord can be analytically
evaluated for two-mode Gaussian states, where one mode is probed through
(single-mode) Gaussian POVMs. The latter can be written in general as
$$\Pi_B(\eta) = \pi^{-1} D_B (\eta) \varrho_M D_B^\dag (\eta)$$ where 
$D_B
(\eta) = \exp(\eta b^\dag- \eta^\ast b)$  is the displacement 
operator, and $\varrho_M$ is a single-mode Gaussian state with zero mean and
covariance matrix $\sigma_M = \left(\begin{array}{cc} \alpha & \gamma \\
\gamma & \beta \end{array} \right) $.
Two-mode Gaussian states can be characterized by their covariance matrix
$ \sigma_{AB} = \left(\begin{array}{cc} A & C \\ C^T & B \end{array}
\right) $. By means of local unitaries that preserve the Gaussian
character of the state, i.e. local symplectic operations, $\sigma_{AB}$
may be brought to the so-called standard form, i.e. $A = \mbox{diag}(a,
a)$, $B = \mbox{diag}(b, b)$, $C = \mbox{diag}(c_1 , c_2 )$. The
quantities $I_1 = \det A$, $I_2 = \det B$, $I_3 = \det C$, $I_4 = \det
\sigma_{AB} $ are left unchanged by the transformations, and are thus
referred to as symplectic invariants. The local invariance of the
discord has therefore two main consequences. On the one hand,
correlation measures may be written in terms of symplectic invariants
only. On the other hand, we can restrict to states with $\sigma$ already
in the standard form.  Before the measurement we have
\bae
S (\varrho_{AB}) = h (d_+) + h (d_- ) , \\
S(\varrho_A)=h(\sqrt{I_1}), \ S(\varrho_B)=h(\sqrt{I_2})
\eae
where $h[x] = (x + 1/2) \log(x + 1/2) - (x - 1/2) \log(x - 1/2) $ and
$d_{\pm}$ are the symplectic eigenvalues of $\varrho_{AB}$ expressed by
$d_{\pm}^2 = 1/2 [ \Delta \pm \sqrt{\Delta^2 - 4 I_4}$, $\Delta = I_1 +
I_2 + 2 I_3$.  After the measurement, the (conditional) 
post-measurement state of mode $A$ is a
Gaussian state with covariance matrix $\sigma_P$ that is independent of
the measurement outcome and is given by the Schur complement $ \sigma_P
= A - C(B + \sigma_M)^{-1} C^T $. 
The Gaussian discord is therefore expressed by
\begin{eqnarray}
D^{\mathcal{G}}(A:B) = h(\sqrt{I_2}) - h(d_{-}) - h(d_{+} ) \nonumber \\
+ \mbox{min}_{\sigma_M} h( \det\sqrt{\sigma_P})
\end{eqnarray}
where we use two key properties: i) the entropy of a Gaussian state
depends only on the covariance matrix, and ii) the covariance matrix
$\sigma_P$ of the conditional state does not depend on the outcome of
the measurement.  The minimization over $\sigma_M$ can be done
analytically. For the relevant case of states with $C = \mbox{diag}(c ,
\pm c )$, including STS and MTS (see below), the minimum is obtained for
$ \alpha = \beta = 1/2, \gamma = 0 $ i.e. when the covariance matrix of
the measurement is the identity. This corresponds to the coherent state
POVM, i.e. to the joint measurement of canonical operators, say position
and momentum, which may be realized on the radiation field by means of
heterodyne detection. For {\em separable} states the Gaussian discord grows
with the total energy of the state and it is bounded, $D \leq 1$;
furthermore, we have $D=0$ iff the Gaussian state is in product form
$\varrho_{AB} = \varrho_A \otimes \varrho_B $. 
\section{non Gaussian discord}  
\label{Sec : NonGaussian correlations}
In this work we consider Gaussian states, and ask whether non-Gaussian
measurements can allow for a better extraction of information than
Gaussian ones, hence leading to lower values of discord. 
\par 
The optimality of Gaussian measurements has been already proven for a
special case~\cite{AdessoDatta}: that of two-mode Gaussian states having
one vacuum normal mode. Indeed any bipartite state $\varrho_{AB}$ can be
purified, $\varrho_{AB}$ $\Longrightarrow$ $|\psi\rangle_{ABC}$; then,
the Koashi-Winter relation~\cite{Koashi}, \begin{equation}
D(A:B)=E_f(A:C)+S(\varrho_B)-S(\varrho_{AB})  \label{Koashi}
\end{equation} relates the quantum discord $D$ and the entanglement of
formation $E_f$ of reduced states $\varrho_{AB}$ and $\varrho_{AC}$
respectively. Given a (mixed) two-mode Gaussian state $\varrho_{AB}$,
there exists a Gaussian purification $|\psi\rangle_{ABC}$. In general,
the purification of $\varrho_{AB}$ requires two additional modes, so
that $\varrho_{AC}$ is a three-mode Gaussian state. In the special case
when one normal mode is the vacuum, the purification requires one mode
only. In this case, $\varrho_{AC}$ represents a two-mode Gaussian state
and $E_f(A:C)$ can be evaluated~\cite{EOF}. From $E_f(A:C)$, by means of
Eq. (\cite{Koashi}), one can obtain $D(A:B)$ (the exact discord) and a
comparison with $D^{\mathcal{G}}(A:B)$ proves that
$D(A:B)=D^{\mathcal{G}}(A:B)$. \par 
In the general case, there is no straightforward analytical way to prove that 
Gaussian discord is optimal. Therefore, we perform a numerical study. Since 
taking into account the most general set of non-Gaussian measurements is an 
extremely challenging task, one can rather focus on a restricted subset. 
We choose to focus on a class of measurements that are realizable with
current or foreseable quantum optical technology. These are the
the projective POVMs, $\Pi = \{ \Pi_n\}$, represented by the following 
orthogonal measurement bases:
\be
\Pi_n=D(\alpha)S(r)\ketbra{n}{n}S(r)^\dagger 
D(\alpha)^\dagger, \,\,\,\, n=0,\cdots,\infty  \label{Eq: class}
\ee
where $S(r)=\exp{(-r^*\frac{a^2}{2}-r\frac{(a^\dagger)^2}{2})}$ and
$D(\alpha)=\exp(\alpha a^\dag - \alpha^* a)$ are respectively the
single-mode squeezing and displacement operators~\cite{DO}. The set of
projectors in (\ref{Eq: class}) is a POVM for any fixed value of 
$\alpha$ and $r$.  If $\alpha=r=0$ we
have the spectral measure of the number operator, describing ideal 
photon counting $\Pi_n = \ket{n} \bra{n}$. If
$\alpha > 0, r=0$ we are projecting onto displaced number
states~\cite{DisplacedNumber}, if $\alpha=0, r > 0 $ onto squeezed
number states~\cite{SNS1,SNS2,SNS3,SNS4}. While more general non Gaussian
measurements are in principle possible, the class (\ref{Eq: class})
encompasses most of the measurements that can be realistically accessed.
\par
In the following, we will evaluate the
non-Gaussian quantum discord defined by
\begin{align}
D^{\mathcal{NG}}(A:B)  = h(\sqrt{I_2})  -  h(d_{-}) - h(d_{+} ) 
+  S^{\Pi,\mathcal{NG}}(A|B)   \label{Eq: nonGdiscdef}
\end{align}
where the non-Gaussian measurements are given by Eq. (\ref{Eq: class}) above.
For the non-Gaussian conditional entropy we have
\begin{align}
S^{\Pi,\mathcal{NG}}(A|B) &= \sum_n p_n S(\varrho_{A,n})\,, \notag \\ 
  \varrho_{A,n} &= \frac1{p_n}\mbox{Tr}_B [\Pi_n \varrho_{AB} \Pi_n]\,, \notag \\ 
  p_n &= \mbox{Tr}_{AB} [\Pi_n \varrho_{AB} \Pi_n]
\end{align}
In the following we consider two classes of Gaussian
states in order to assess the performances of the above 
measurements. These are the two-mode squeezed thermal states 
(STS)~\cite{STSteo,STSexp1,STSexp2}:
\be
\varrho =S(\lambda)\nu_1(N_1)\otimes\nu_1(N_2)S(\lambda)^\dagger
\label{Eq.: rhoSTS}
\ee
and the two-mode mixed thermal states (MTS) \cite{MTS}
\be
\rho =U(\phi)\nu_1(N_1)\otimes\nu_1(N_2)U(\phi)^\dagger
\label{Eq.: rhoMTS}
\ee
where $\nu_i(N_i)$ are 1-mode thermal states with thermal photon number
$N_i$; $S(\lambda)=\exp\{\lambda (a_1^\dag a_2^\dag - a_1 a_2) \}$ is
the two-mode squeezing operator (usually realized on optical modes
through parametric down-conversion in a nonlinear crystal); and
$U(\phi)=\exp\{\phi( a_1^\dag a_2 - a_1 a_2^\dag)\}$ is the two-mode
mixing operator (usually realized on optical modes through a beam
splitter).  
\par 
In particular, in the following we will focus on the simplest case of
symmetric STS with $N_1=N_2 \in [10^{-5},1]$ $\lambda \in [0,0.5]$.  As
for MTS, we cannot consider the symmetric case (since if $N_1=N_2$ then
the mutual information vanishes and there are no correlations in the
system), therefore we consider the unbalanced case and focus on $\phi
\in [0,\pi/2]$ and $N_1,N_2 \in [10^{-5},1]$.
\section{Number basis} \label{Sec : number basis}

Let $\Pi_n=\ketbra{n}{n}$. In this case, the post-measurement state is
\be
\varrho^A_n\otimes\ketbra{n}{n}=\left(\sum_{h,k}\varrho_{(h,k),(n,n)}\ketbra{h}{k}\right)\otimes\ketbra{n}{n}
\ee
and we have the following expression for the density matrix elements
\be
\varrho_{(h,k),(n,n)}=\sum_{s,t} p^{th}_s (N_1) p^{th}_t (N_2) O_{hn}(st)O_{kn}^*(st)
\label{Eq.: rhoApostnumber}
\ee
where $ p^{th}_s (N) = N^s \ (1+N)^{-(s+1)}  $ and $O_{hn}(st) = \bra{hn} O \ket{st} $ with $O=S(\lambda),U(\phi)$ for STS and MTS respectively.
The post-measurement state $\varrho^A_n$ is diagonal (see appendix \ref{Sec. : postdiagonal}),
\begin{equation}
\bra{h} \varrho_n^A \ket{k} = \delta_{hk} \ \varrho_{(h,h),(n,n)}
\end{equation}
As a consequence, the entropy of the post-measurement state can be expressed as: $S(\varrho^A_n)=H(\{ \varrho_{(h,h),(n,n)}\})=H(\vec{p}(A|B=n))$
where $H$ is the Shannon entropy of the conditional probability $\vec{p}(A|B=n)=(p(0,n),p(1,n),\cdots)/p_n$, and therefore the overall conditional entropy  can be simply expressed in terms of the photon number statistics:
\bae
S(A|\{\Pi_n\})&=&\sum_n p_n H(\vec{p}(A|B=n))= \nonumber \\&=&
H(\vec{p}(A,B))-H(\vec{p}(B))
\label{Eq.: CondEntSTSnumber}
\eae
with $\vec{p}(A,B)=\{p(A=n,B=m)\}$ and $\vec{p}(B)=\{p(B=n)\}$.
In view of this relation, the only elements of the number basis
representation of the density matrix $\varrho$ that are needed are the
diagonal ones, i.e. one has to determine the photon number statistics
for the two-mode STS or MTS state. The required matrix elements can be
obtained in terms of the elements of the two-mode squeezing and mixing
operators (see appendix \ref{Sec. : postdiagonal}).  One has of course
to define a cutoff on the dimension of the density matrix. This can be
done upon requiring that the error on the trace of each state considered
be sufficiently small: $\epsilon_{err}=1-\mbox{Tr}\varrho\le 10^{-3}$.
\begin{figure}[t]
\centering
\includegraphics[width=0.4\textwidth]{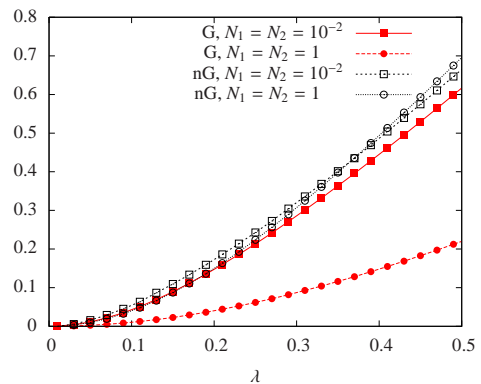}
\caption{ Gaussian and non-Gaussian quantum discord for STS as a function of $\lambda$, for different values of $N_1=N_2$}
\label{Fig: number basis qdisc}
\end{figure}

\begin{figure}[h]
\centering
\includegraphics[width=0.4\textwidth]{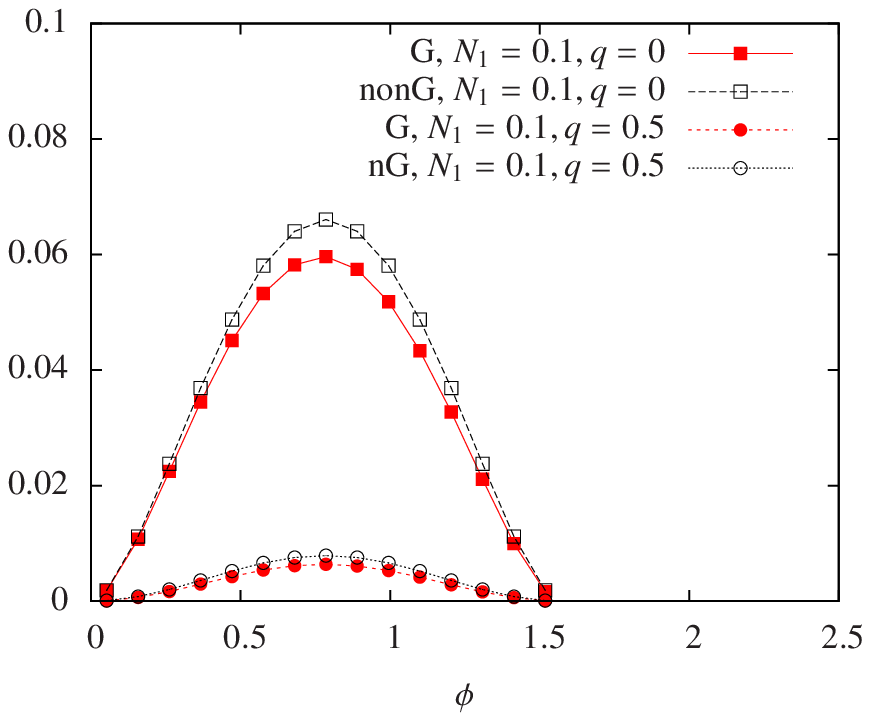}
\includegraphics[width=0.4\textwidth]{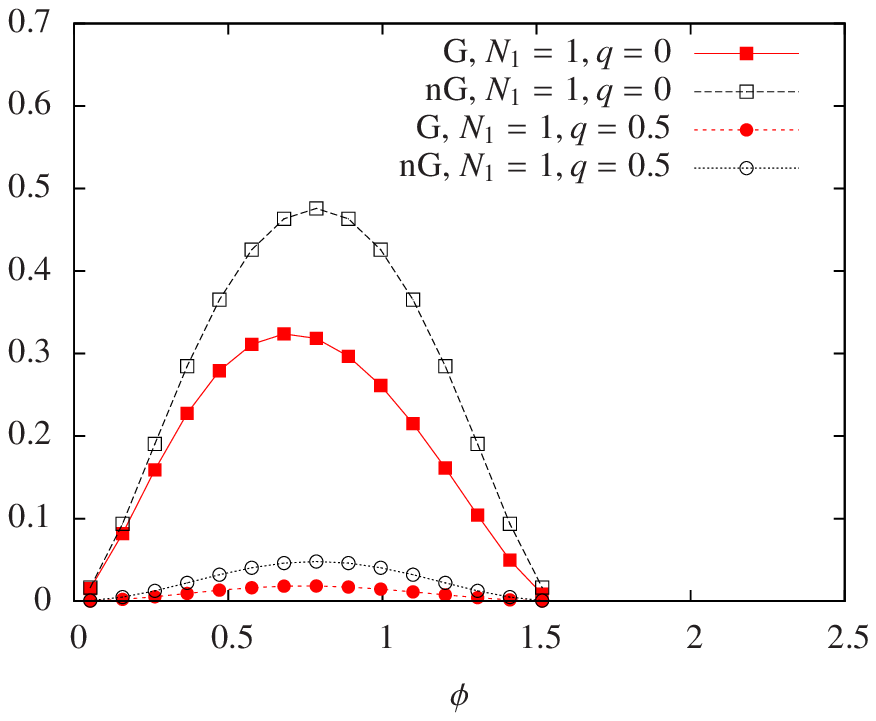}
\caption{Gaussian and non-Gaussian quantum discord for MTS states as a function of $\phi$ for different values of $N_1$ and $q=N_2/N_1$}
\label{Fig: number basis mts}
\end{figure}

We have compared Gaussian and non-Gaussian quantum discord (with the non-Gaussian measurements corresponding to photon number measurements) for STS and MTS states with a wide range of squeezing, mixing and thermal parameters. In Fig. \ref{Fig: number basis qdisc} we show results for STS with varying $\lambda$ and $N_1=N_2=10^{-2}$, $N_1=N_2=1$. The key result is that the non-Gaussian quantum discord is always greater than its Gaussian counterpart for all values of $N_1$ and $\lambda$. The gap grows with increasing $N_1$ and $\lambda$.
In Fig.~\ref{Fig: number basis mts} we show results for MTS $N_1=\{0.1, 1\}$ and $q=N_2/N_1=\{0,0.1,0.4,0.5\}$. Also in this case, the non-Gaussian discord is always higher than the Gaussian one.\\
Both results indicate that the Gaussian (heterodyne) measurement is optimal for STS and MTS states, at least compared to photon counting, in the sense that it allows for a better extraction of information on mode $A$ by a measurement on mode $B$.

\section{Squeezed Number basis} \label{Sec : squeezed number basis}

\begin{figure}[bt]
\centering
\includegraphics[width=0.4\textwidth]{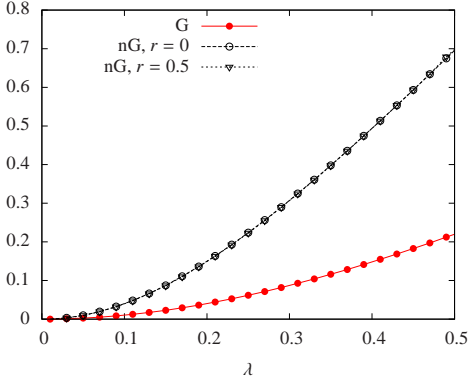}
\caption{Gaussian and non-Gaussian quantum discord for STS with $N_1=1$ as a function of $\lambda$ and for different values of local squeezing $r$}
\label{Fig: squeezed number basis qdisc}
\end{figure}

\begin{figure}[h]
\centering
\includegraphics[width=0.4\textwidth]{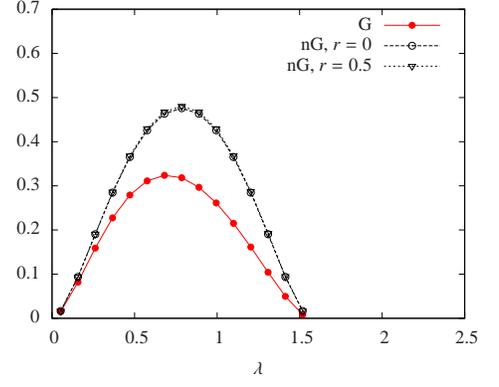}
\caption{Gaussian and non-Gaussian quantum discord for MTS states for $N_1=1$, $N_2=0$ as a function of $\phi$  and for different values of local squeezing $r$}
\label{Fig: squeezed number basis mts}
\end{figure}

We now analyze the case of non-Gaussian measurements represented by the squeezed number basis $\ketbra{n_r}{n_r}=S(r)\ketbra{n}{n}S(r)^\dagger$, where $S(r)=\exp{(-r^*\frac{a^2}{2}-r\frac{(a^\dagger)^2}{2})}$ is the single mode squeezing operator.
A local measurement in the squeezed number basis is equivalent to a measurement in the number basis, performed on a locally squeezed state. In formulas, the probability of measuring $n_r$ on one subsystem when the state is the $\varrho$ is
\bae
p_\varrho(n_r)&=&\Tr(\openone\otimes\ketbra{n_r}{n_r}\varrho)=\Tr(\openone\otimes\ketbra{n}{n}S^\dagger(r)\varrho S(r))= \nonumber
\\&=& \Tr(\openone\otimes\ketbra{n}{n}\varrho^r)=p_{\varrho_r}(n)
\label{Eq.: probSquezeednumber}
\eae
i.e., is equal to the probability of measuring $n$ on the locally squeezed state $\varrho_r = S(r) \varrho S(r)^\dag$, and the relative post-measurement state is
\bae
\varrho_{n_r}^A&=&\Tr_B[\openone\otimes\ketbra{n_r}{n_r}\varrho\openone\otimes\ketbra{n_r}{n_r}]/p_\varrho(n_r)=\nonumber \\&=&
\Tr_B[\openone\otimes\ketbra{n}{n}\varrho_r\openone\otimes\ketbra{n}{n}]/p_\varrho^r(n)=
\varrho_{r_{n}}^A
\label{Eq.: rhoApostSqueezednumber}
\eae
The general idea is that measurements on a state $\varrho$ in a basis
that is obtained by performing a unitary (Gaussian) operation $V$ on the
number basis $\ketbra{n}{n}$ can be represented as measurements on the
number basis of a modified state $\varrho_V = V \varrho V^\dag$ on which
the local unitary operation acts.\\ In the case of the squeezed number
basis, the post-measurement state is not diagonal, therefore the
reasoning leading to Eq. (\ref{Eq.: CondEntSTSnumber}) does not hold.
The post-measurement state matrix elements
$(\varrho_{r_{n}}^A)_{h,k}=\varrho_{(h,k),(n,n)} $ can be obtained
directly by evaluating the expression (\ref{Eq.: rhoApostnumber}) where
now the expression $O_{hk}(st)=\bra{h k}O\ket{st}$ (where
$O=S(\lambda),U(\phi)$) must be substituted with $O_{hk}'(st)=\bra{h
k}S(r)O\ket{st} = \sum_q \bra{k} S(r) \ket{q} \bra{hq} O\ket{st} $, and
the elements of the single mode squeezing operator are given in
\cite{SingleModeSqueezing} (eq. 20) or in
\cite{SingleModeSqueezingKnight} (eq. 5.1).\\ We have evaluated the
Gaussian and non-Gaussian quantum discord for STS and MTS states with a
wide range of two-mode squeezing and thermal parameters. Non-Gaussian
measurements are done in the squeezed photon number basis, $\Pi_n =
S(r)\ket{n} \bra{n} S(r)^\dag$ with variable $r \in [0,0.5]$.  The
effect of local squeezing on non-Gaussian quantum discord is negligible
in the whole parameter range under consideration: we compare the
non-Gaussian discord for different values of $r$ and find that all
curves collapse. This can be seen in fig. \ref{Fig: squeezed number
basis qdisc} and fig. \ref{Fig: squeezed number basis mts} where plot
the behavior for $N_1=N_2=0.01$ (STS) and $N_1=1, N_2=0$ (MTS). The same
behavior is observed in the whole parameter range under investigation.
We have verified numerically that the post-measurement states of mode A
$\varrho_{r_{n}}^A$ are not equal as $r$ varies (i.e., the
post-measurement states corresponding to measurement result $n_r$ change
with $r$), yet the sum $\sum_n p_n S(\varrho_{r_{n}}^A)$ is equal for
all values of $r$ under investigation. Therefore, the squeezing in the
measurement basis has no effect on the discord, at least for the values
of squeezing considered: in particular, it cannot afford a deeper
minimization than that obtained without local squeezing. This indicates
that the heterodyne measurement remains optimal also with respect to
measurement in the squeezed number basis.
\section{Displaced Number basis}  \label{Sec : displaced number basis}

We finally analyze the case of non-Gaussian measurements represented by the displaced number basis $\ketbra{n_\alpha}{n_\alpha}=D(\alpha)\ketbra{n}{n}D(\alpha)^\dag$, where $D(\alpha)=\exp(\alpha a^\dag -\alpha^* a)$ is the single mode displacement operator. According to the general considerations above, a local measurement in the displaced number basis is equivalent to a measurement in the number basis, performed on a locally displaced state $\varrho_\alpha$.
As in the case of the squeezed number basis, the post-measurement state is not diagonal and we need all matrix elements $(\varrho_{\alpha_{n}}^A)_{h,k}=\varrho_{(h,k),(n,n)} $. They can
be obtained directly by evaluating the expression (\ref{Eq.: rhoApostnumber}) where the expression $O_{hk}(st)=\bra{h k}S(\lambda)\ket{st}$ (where $O=S(\lambda),U(\phi)$) must be substituted with $O_{hk}'(st)=\bra{h k}D(\alpha)O\ket{st}=\sum_q \bra{k} D(\alpha)\ket{q} \bra{hq} O\ket{st}$, and the elements of the single mode displacement operator are given in~\cite{Parisbook} (eq. 1.46).

\begin{figure}[bt]
\centering
\includegraphics[width=0.4\textwidth]{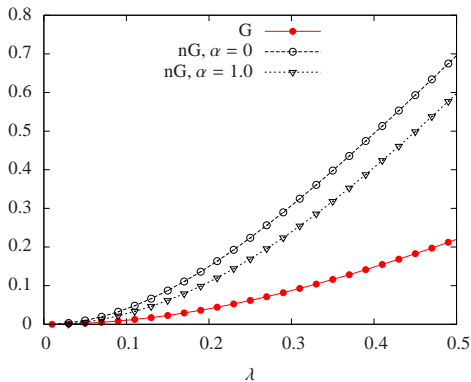}
\caption{Gaussian and non-Gaussian quantum discord for STS with $N_1=1$ as a function of $\lambda$ and for different values of local displacement $\alpha$}
\label{Fig: displaced number basis qdisc}
\end{figure}

\begin{figure}[h]
\centering
\includegraphics[width=0.4\textwidth]{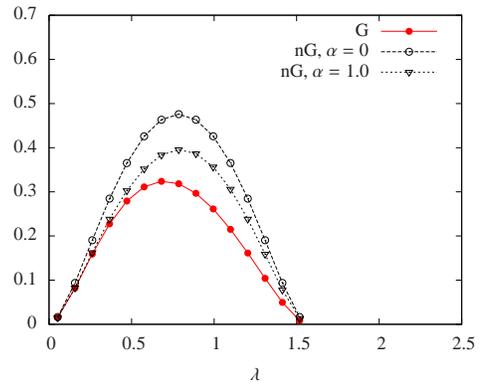}
\caption{Gaussian and non-Gaussian quantum discord for MTS states for $N_1=1$, $N_2=0$ as a function of $\phi$  and for different values of local displacement $\alpha$}
\label{Fig: displaced number basis mts}
\end{figure}

The evaluation of the non-Gaussian quantum discord can be simplified by first noticing that one can consider real values of $\alpha$ only. Indeed, the quantum discord only depends on the modulus $|\alpha|$. This is shown in detail in the appendix~\ref{Sec : phase irrelevant}, by using the characteristic function formalism. Consider $\varrho_{n_\alpha}^A $,
the post-measurement state of mode $A$ after measurement result $n_{\alpha}$ is obtained on $B$. If we change the phase of $\alpha$,
$\alpha \rightarrow \alpha' \equiv e^{ i \theta} \alpha $  we find that
\be
\varrho_{n_{\alpha'}}^A  = U \varrho_{n_\alpha}^A  U^\dag
\ee
where $U$ is a unitary operation corresponding to a simple quadrature rotation
\begin{equation}
a_1 \to a_1 e^{i \theta}  \quad a_1^\dag \to a_1^\dag e^{-i \theta}
\end{equation}
 Therefore, we have $\varrho_{\alpha'_n}^A \neq \varrho_{\alpha_n}^A$, but
$\varrho_{\alpha'_n}^A$ and $\varrho_{\alpha_n}^A$ have the same spectrum, since they are related by a unitary.
Therefore, the entropy of the reduced post-measurement state $\varrho_{\alpha}^A$ does not depend on the phase of $\alpha$ but just on $|\alpha|$.  If follows that the non-Gaussian quantum discord of $\varrho_{\alpha}$ does not depend on the phase of $\alpha$. \\
We have evaluated the Gaussian and non-Gaussian quantum discord for STS and MTS states with a wide range of two-mode squeezing and thermal parameters.  
Non-Gaussian measurements are done in the displaced photon number basis, $\Pi_n = D(\alpha)\ket{n} \bra{n} D(\alpha)^\dag$ with variable $\alpha \in [0,2.5]$.
In fig.~\ref{Fig: displaced number basis qdisc} and fig.~\ref{Fig: displaced number basis mts} we plot the Gaussian and non-Gaussian quantum discord. We see that greater displacements lead to lower values of the non-Gaussian quantum discord, but the decrease is insufficient to match the Gaussian quantum discord, which remains optimal. However, the non-Gaussian quantum discord approximates the Gaussian one as $\alpha \to \infty$. This is analytically proven below in the appendix~\ref{Sec : growing alpha}. There we find that for both STS and MTS
\begin{eqnarray}
& \ & \varrho_{\alpha_n}^A  \to \varrho_{\alpha_0}^A  
\qquad \mbox{as} \quad \alpha \to \infty \label{Eq: limit}
\end{eqnarray}
i.e, the conditional states $\varrho_{\alpha_n}^A$ becomes independent
of $n$ and equal to the $n=0$ result. As a consequence, the conditional
entropy in the displaced number basis is equal to the entropy of the
post-measurement state for any measurement result, and, in particular,
for $n=0$: \begin{equation}
S^{\Pi,\mathcal{NG}} (A|B) = \sum_n   
p_n S(\varrho_{\alpha_n}^A) \to S(\varrho_{\alpha_0}^A) 
\label{Eq: conditonalnonG} \quad \mbox{as} \quad \alpha \to \infty
\end{equation}
But $\varrho_{\alpha_0}^A $ is just the post-measurement state we obtain
after a heterodyne detection on mode $B$ (equal for all measurement
result modulo a phase space translation which
is irrelevant as for the entropy). Therefore, we also have $S^{G} (A|B)
= S(\varrho_{\alpha_0}^A) $ and the non Gaussian discord
$D^{\mathcal{NG}}(A:B)$ in the displaced number basis tends to the
Gaussian discord $D^{\mathcal{G}}(A:B)$ as $\alpha \to \infty$. \par 
Actually, we cannot prove that the $D^{\mathcal{NG}}(A:B)$ is lower
bounded by  $D^{\mathcal{G}}(A:B)$, and we cannot rule out the
possibility that $D^{\mathcal{NG}}(A:B) <  D^{\mathcal{G}}(A:B)$  for
intermediate values of $\alpha$. However, our numerical data do not
support this possibility since we never observe $D^{\mathcal{NG}}(A:B) <
D^{\mathcal{G}}(A:B)$ and we expect that $D^{\mathcal{NG}}(A:B) \to
D^{\mathcal{G}}(A:B)$ from above as $\alpha \to \infty$. \\ In
conclusion, we have analytical and numerical evidence that the
heterodyne measurement remains optimal also with respect to measurement
in the displaced number basis.
\section{Geometric discord}  \label{Sec : Geometric}
In this section, we briefly consider the recently introduced measure of
geometric discord and compare results with those obtained for the
quantum discord. The geometric discord~\cite{Dakic} is defined as
\begin{equation}
D_G(A:B) = \min_{\{\chi_{AB} \in \mathcal{C} \}} || \varrho_{AB} - \chi_{AB} ||_2
\end{equation}
and it measures the distance of a state from the set $\mathcal{C}$ of
quantum-classical states where $||A||_2 = \mbox{Tr}[A^\dag A]$ is the
Hilbert-Schmidt distance. Clearly $D_G = 0$ iff $D=0$, since both
measures vanish on the set of classically correlated states. In
particular, it has been be proven that $D_G$ can be seen a measure of
the discrepancy between a state before and after a local measurement on
subsystem $B$~\cite{Luo}: 
\begin{equation}
D_G(A:B)=\min_{\{\Pi \in POVM \}} || \varrho_{AB} - \varrho_{AB}^\Pi ||_2    \label{Eq.: geomdisc}
\end{equation}
where the unconditional post-measurement state is given by
$\varrho_{AB}^\Pi = \sum_x M_B^x \varrho_{AB} M_B^{x \dag} $.  Notice
that $D_G$ and $D$ are not monotonic functions of one
another and the relation between them is still an open question.
However, in many cases $D_{G}$ is much simpler to evaluate than $D$.  
\par
Analogous to the case of Gaussian discord, a Gaussian version of the
geometric discord can be defined by restricting to  Gaussian
measurements~\cite{GaussianGeom}. Again, it can be analytically computed
for two-mode Gaussian states. With the same reasoning of sec. \ref{Sec :
correlations} one easily obtains 
\be
D_G^{\mathcal{G}} (\varrho_{AB} ) = 
\mbox{min}_{\sigma_M} \mbox{Tr}[(\varrho_{AB}-\varrho_P \otimes \varrho_M)^2]
\ee
Exploiting the property that $\mbox{Tr}[\varrho_1 \varrho_2] = 1/\det[(\sigma_1 + \sigma_2)/2],$ for any two Gaussian states $\varrho_1$ and $\varrho_2$,
\bae
D_G (A:B) = \mbox{min}_{\sigma_M} \{ 1/\sqrt{\det\sigma_{AB}} +   \\
\nonumber + 1/\sqrt{ \det(\sigma_P \oplus \sigma_M )} - 2/\sqrt{\det[(\sigma_{AB} + \sigma_P \oplus \sigma_M)/2]}  \}
\eae
For for the relevant case of STS and MTS, the minimum is obtained with
the $\sigma_M$ elements given by $ \alpha = \beta = \frac{\sqrt{ab}
(\sqrt{4 ab - 3 c^2} + \sqrt{ab})} {3a}$, $\gamma = 0 $. The least
disturbing Gaussian POVM for STS, according to the Hilbert-Schmidt
distance, is thus a (noisy) heterodyne detection, a result which is
analogous to what found in the case of quantum discord. If one
constrains the mean energy per mode, the Gaussian quantum
discord gives upper and lower bounds to the Gaussian geometric discord.
In absence of such a provision, the geometric discord can vanish for
arbitrarily strongly nonclassical (entangled) Gaussian states, as a
consequence of the geometry of CV state spaces. \par Also in this case, we 
may consider
non-Gaussian measurements and evaluate a non-Gaussian geometric discord:
\be
D_G^{\mathcal{NG}} (A:B) = \mbox{Tr}[(\varrho_{AB}-\varrho_{AB}^\Pi)^2]  \label{Eq: nonGgeomdef}
\ee
For measurement in the number basis, we can easily obtain
\bae
D_G^{\mathcal{NG}} =   \mu(\varrho) + \sum_{n pq} |\bra{p n }\varrho \ket{q n}|^2   \label{Eq.: nonGausgeom}
\eae
where $\mu(\varrho)=\frac{1}{4 \sqrt{\det(\sigma)}}$ is the (Gaussian) state purity~\cite{Parisbook}. 
In the case of measurements in the squeezed or displaced number basis, we have to use $\varrho_r$ and $\varrho_\alpha$
instead of $\varrho$ in Eq. (\ref{Eq.: nonGausgeom}). In general, in order to compute the geometric discord we need to compute matrix elements, and we use the same numerical methods described above. \\


\begin{figure}[bt]
\centering
\includegraphics[width=0.4\textwidth]{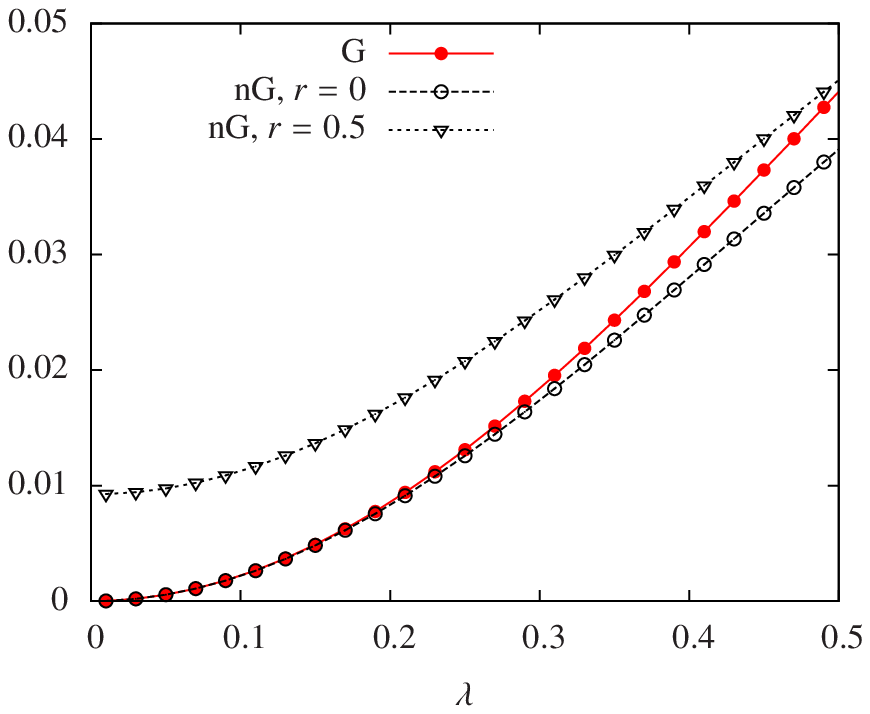}
\includegraphics[width=0.4\textwidth]{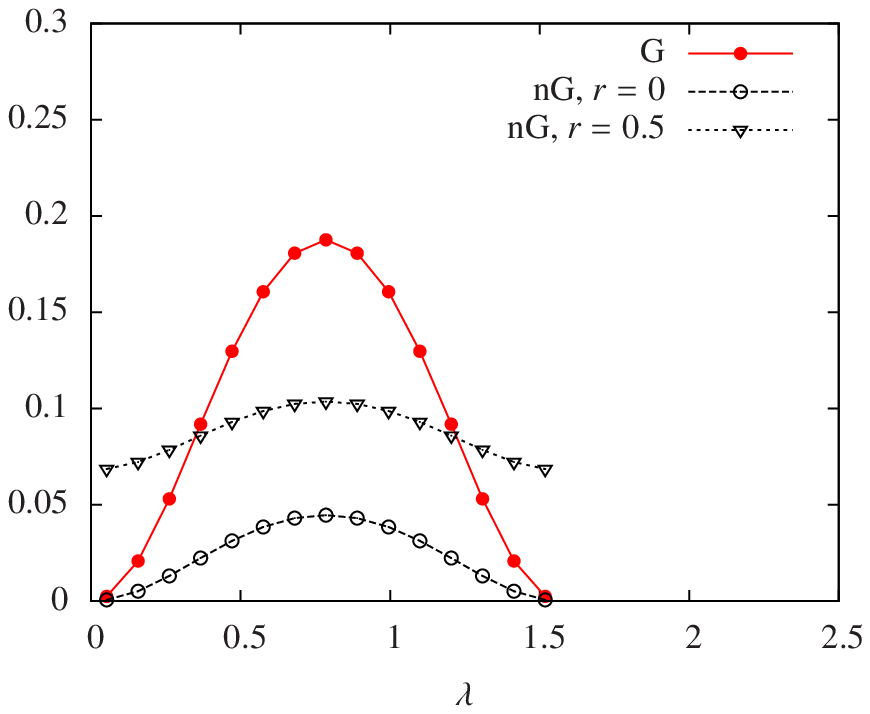}
\caption{ (Top) Gaussian and non-Gaussian geometric discord for STS with $N_1=1$ as a function of $\lambda$ and for different values of local squeezing $r$; (Bottom) Gaussian and non-Gaussian geometric discord for MTS states for $N_1=1$, $N_2=0$ as a function of $\phi$  and for different values of local squeezing $r$}
\label{Fig: squeezed number basis geomdisc}
\end{figure}

\begin{figure}[bt]
\centering
\includegraphics[width=0.4\textwidth]{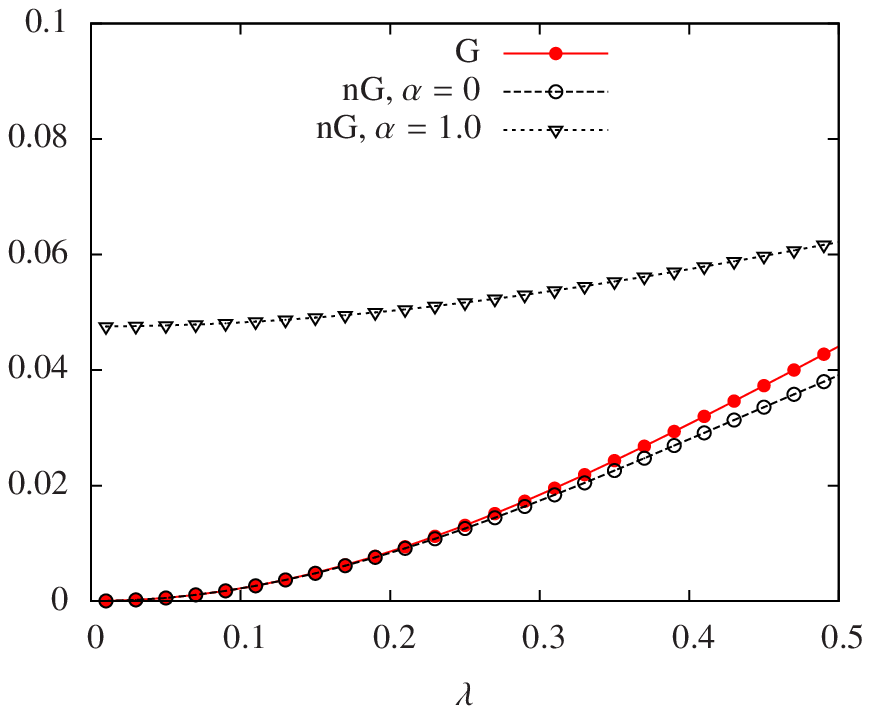}
\includegraphics[width=0.4\textwidth]{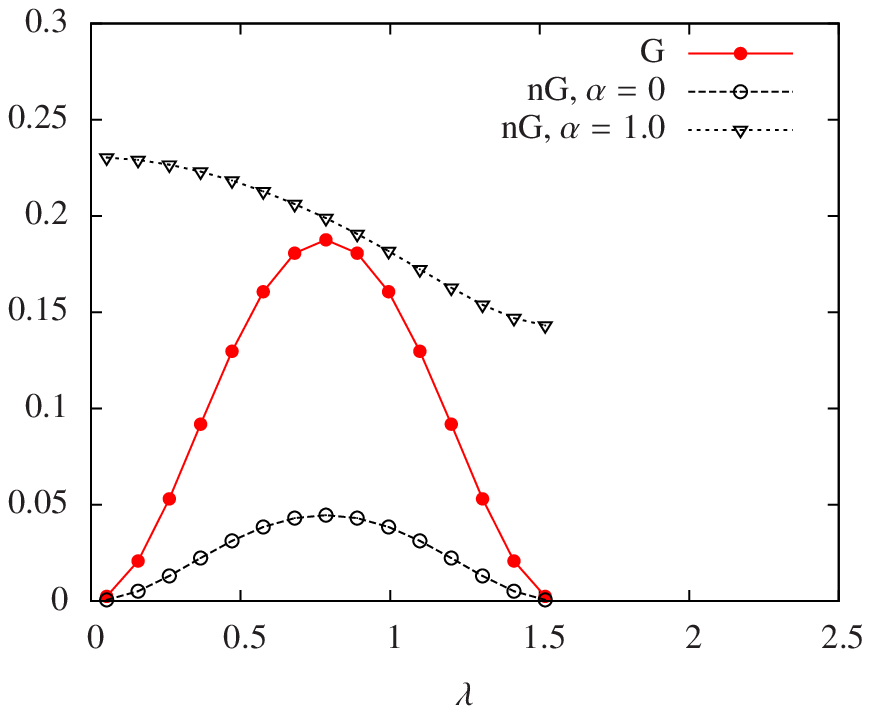}
\caption{(Top) Gaussian and non-Gaussian geometric discord for STS with $N_1=1$ as a function of $\lambda$ and for different values of local displacement $\alpha$ (Bottom) Gaussian and non-Gaussian geometric discord for MTS states for $N_1=1$, $N_2=0$ as a function of $\phi$  and for different values of local displacement $\alpha$}
\label{Fig: displaced number basis geomdisc}
\end{figure}

\subsection{Results}. 
We have compared the Gaussian and non-Gaussian geometric discord for STS
and MTS in a wide range of parameters. We have considered measurements
in the number, squeezed number and displaced number basis for the same
values of the parameters given in the preceding sections. Results are
plotted in Figs.~\ref{Fig: squeezed number basis geomdisc} and \ref{Fig:
displaced number basis geomdisc}. In general, at variance with the
results for quantum discord, we find that non-Gaussian measurements can
provide lower values of geometric discord than Gaussian ones. 
Among the class of non-Gaussian measurements we have considered, 
the optimal one is provided by the number basis, 
which gives values of geometric discord that are always lower than those
given by the optimal Gaussian measurement.  The non-Gaussian geometric
discord increases for increasing $r$ and
$\alpha$, and it can become greater than its Gaussian counterpart. These
results are very different from the quantum discord case: on one hand,
the (non-Gaussian) geometric discord is substantially affected by the
local squeezing; on the other hand, it does not approach the Gaussian
one when the displacement $\alpha \to \infty$, but it grows
monotonically. Indeed if we increase the squeezing or displacement in
the measurement basis, the post-measurement state is more distant (in
Hilbert-Schmidt norm) from the original one. As already noticed,
performing the measurement is the squeezed (displaced) number basis in
equivalent to first squeezing (displacing) the state and then measuring
it in the number basis. The local squeezing and displacement have the
effect of increasing the energy of the state, shifting the photon number
distribution $P(B=n)$ towards greater values of $n$.  This causes the
overlap between the post measurement state and the original state to
decrease, and therefore their distance to increase. \par 
Let us futher comment on the difference between the quantum discord and the geometric discord cases. 
Quantum discord and geometric discord both vanish for classical states,
but are not monotonic functions of one another, and thus they are truly
different quantities. The geometric discord, based on the
Hilbert-Schmidt distance, is a geometric measure of how much a state is
perturbed by a local measurement, while quantum discord assesses to
which extent correlations are affected by a local measurement.  While
for the quantum discord well-defined operational and informational
interpretations can be found~\cite{Gu,Datta2}, for the geometric discord
the situation is more problematic.  Indeed, one can design protocols in
which the geometric discord can in some cases be related to the protocols'
performances \cite{RSP, Tufarelli}; however, recent discussions
~\cite{Piani}, show that, as consequence of the noninvariance of the
Hilbert-Schmidt norm under quantum evolutions, it is difficult to find a
conclusive argument about the relevance of geometric discord as a
measure of quantumness of correlations.  Our data show that non-Gaussian
measurements can yield optimal values of the geometric discord, contrary
to the case of quantum discord. Hence, the behavior of quantum discord and
geometric discord with respect to different types of measurements is
different. This is a further indication that the geometric discord
cannot be used as a good benchmark for the quantum discord and that the
degree of quantumness measured, if any, by such a quantity has a
fundamentally different nature.
\section{Discussion and conclusions} \label{Sec. : conclusions}
The definition of discord involves an optimization over all possible
local measurements (POVMs) on one of the subsystems of a bipartite
composite quantum system.  In the realm of continuous variables (CV),
initial research efforts on quantum discord restricted the minimization
to the set of (one-mode) Gaussian measurements.  
\par
In this work we have
investigated CV quantum discord beyond this restriction.  We have
focused on Gaussian states, asking whether Gaussian measurements are
optimal in this case, i.e., whether the Gaussian discord is the true
discord for Gaussian states. While a positive answer to this question had
already been given for the special case of two-mode Gaussian states
having one vacuum normal mode (by means of an analytical argument based
on the Koashi-Winter formula), no general result was available so far.
We have addressed our central question upon considering two large
classes of two-mode Gaussian states, the squeezed thermal states (STS)
and the mixed thermal states (MTS), and allowing for a wide range of
experimentally feasible non-Gaussian measurements based on orthogonal
bases:  the photon number basis, the squeezed number basis, the
displaced number basis. For both STS and MTS states, in the range of
parameters considered, the Gaussian measurements always provide optimal
values of discord compared to the non-Gaussian measurements under
analysis. Local squeezing of the measurement basis has no appreciable
effect on correlations, while local displacement leads to lower values
of the non-Gaussian discord, which approaches the Gaussian one in the
limit of infinite displacement.  
\par
Overall, for the explored range of states and measurements, we have 
evidence that the Gaussian discord is the ultimate quantum discord for Gaussian states. 
We note that the optimality of Gaussian measurements suggested by our
analysis is a property which holds only for Gaussian states. In the case
of non-Gaussian states, e.g., CV Werner states, non-Gaussian
measurements such as photon counting can lead to a better minimization,
as was recently proven in Ref.~\cite{NonGausDisc}. 
\par 
We also have investigated the CV geometric discord~\cite{GaussianGeom},
comparing the Gaussian and non-Gaussian cases. We have shown that the
behavior of geometric discord is completely different from that of
quantum discord.  On one hand, non-Gaussian measurements can lead to
lower values of the geometric discord, the number basis measurement
being the optimal one; on the other hand, the effects of both local
squeezing and displacement are strong and consist in a noteworthy
increase in the non-Gaussian geometric discord. The remarkable
differences between quantum and geometric discord imply that the latter
cannot be used as a benchmark of the former.
\par 
Both in the case of the discord and geometric discord a definite answer
on the optimal measurement minimizing the respective formulas would
require the extension of the set of non-Gaussian measurements to
possibly more exotic ones and the application of those realizable in
actual experiments to a broader class of Gaussian and non-Gaussian
states. While we leave this task for future research, our results on
discord support the conjecture that Gaussian measurements are optimal
for Gaussian states and allow to set, for the class of states analyzed,
a tighter upper bound on the entanglement of formation for $1 \times 2$
modes Gaussian states, via the Koashi-Winter relation.
\appendix

\section{The post-measurement state is diagonal} \label{Sec. : postdiagonal}

We prove that the post-measurement state
\be
\varrho_n^A=\Tr_B[\openone\otimes\ketbra{n}{n}\ \varrho \ \openone\otimes\ketbra{n}{n}]/p_n
\ee
of STS and MTS after local measurement in the number basis is diagonal (here, $p_n=\Tr[\varrho\openone\otimes\ketbra{n}{n}]$). We have indeed:
\bae
\varrho = \sum_{s,t} p^{th}_s(N_1) p^{th}_t (N_2) O \ketbra{st}{st}O^\dagger = \nonumber \\=
\sum_{(h,n),(k,m)}\ketbra{h n}{k m}\left(\sum_{s,t} p^{th}_s p^{th}_t O_{hn}(st)O_{km}^*(st)\right)
\eae
where $ p^{th}_s (N) = N^s \ (1+N)^{-(s+1)}  $ where $O_{hn}(st)=\bra{hn}O\ket{st}$ and $ O_{km}^*(st)=\bra{st}O^\dagger\ket{km}=\bra{km}O\ket{st}^*$, where $O=S(\lambda)$, $O=U(\phi)$ for STS and MTS respectively.
The post measurement states can be written as:
\be
\varrho^A_n\otimes\ketbra{n}{n}=\left(\sum_{h,k}\varrho_{(h,k),(n,n)}\ketbra{h}{k}\right)\otimes\ketbra{n}{n}
\ee
and therefore we need to evaluate the matrix elements
\be
\varrho_{(h,k),(n,n)}=\sum_{s,t} p^{th}_s p^{th}_t O_{hn}(st)O_{kn}^*(st) \label{Eq.: rhoApostnumber2}
\ee
The elements of the two-mode squeezing operator are given in \cite{SNS4} (eq. 22):
\bae
\bra{hn}S(\lambda)\ket{st}=\delta_{t+h,s+n}f^\lambda(h,n,s,t) =\delta_{t+h,s+n} \times  \nonumber \\
\sum_{a=0}^{min(s,t)}\sum_{b=0}^{min(h,n)}(-1)^{a+b} (sech \lambda)^{t+h-a-b-1}\mu^{a-b+h-s} \times \nonumber \\
\frac{(t+h-a-b)![s!t!h!n!]^{1/2}}{a!(t-a)!(s-a)!b!(n-b)!(h-b)!} \qquad \qquad
\label{Eq.: Squeezoperatornumber}
\eae
where $\mu=e^\lambda$, while the elements of the two-mode mixing operator
\bae
\bra{hn}U(\phi)\ket{st} = \delta_{h+n,s+t}\sum_{a=\max\{0, h-t\}}^{\min\{h,s\}} 
A_{a \ h-a}^{s \ t}  \nonumber \\
=\delta_{h+n,s+t}  \sum_{a=\max\{0, h-t\}}^{\min\{h,s\}}
\sqrt{\frac{h!(s+t-h)!}{s!t!}} (-1)^{h-a}  \times \nonumber \\
 \binom{s}{a} \binom{t}{h-a} \sin \phi^{s+h-2a} \cos \phi^{t+2a-h} \qquad \qquad
\label{Eq.: Mixoperatornumber}
 \eae
In order to evaluate (\ref{Eq.: rhoApostnumber2}), we need
$O_{hn}(st)O_{kn}^*(st)$. 
Due to the $\delta$'s appearing in both (\ref{Eq.: Squeezoperatornumber}) and (\ref{Eq.: Mixoperatornumber}),
the following relations must be satisfied:
\bae
t-s&=&n-h\nonumber \\
t-s&=&n-k\nonumber.
\eae
and this implies $h=k$; therefore the post-measurement state is diagonal in the number basis:
\bae
(\varrho_n^A)_{h,k} = \delta_{h,k}\sum_{s,t} p^{th}_s(N_1) p^{th}_t (N_2) |O_{hk}(st)|^2
\eae

\section{Discord does not depend on the phase of displacement} \label{Sec : phase irrelevant}

We show that the (non-Gaussian) discord in the displaced number basis does not depend on the phase of displacement
for STS and MTS. The arguments is best given in the characteristic function representation of the states~cite{Parisbook}
The STS and MTS states have a Gaussian characteristic function $\chi [\varrho](\mathbf{\Lambda}) = \exp(-\frac{1}{2} \mathbf{\Lambda}^T \sigma \mathbf{\Lambda})$ where $\mathbf{\Lambda} = \frac{1}{\sqrt{2}}(\mbox{Re} \lambda_A, \mbox{Im} \lambda_A, \mbox{Re} \lambda_B, \mbox{Im} \lambda_B)$ and the covariance matrix is given by
\begin{equation}
\sigma  = \left( \begin{array}{cc}
A & C \\
C^T & B
\end{array} \right)  =
\left( \begin{array}{cccc}
a & 0 & c & 0 \\
0 & a & 0 & \pm c \\
c & 0 & b & 0 \\
0 & \pm c & 0 & b
\end{array} \right)
\end{equation}
where $\pm c$ is $-c$ in the case of STS and $+c$ in the case of MTS. For STS we have
\begin{equation}
\chi [\varrho](\lambda_A, \lambda_B) = \exp(- a |\lambda_A|^2 - b |\lambda_B|^2 + 2 c \mbox{Re}[\lambda_A \lambda_B] )
\end{equation}
while for MTS the same expression holds upon changing $2 c \mbox{Re}[\lambda_A \lambda_B] \to 2 c \mbox{Re}[\lambda_A^* \lambda_B]$. In the following, we shall carry on the argument for STS, but the MTS case is fully equivalent.
If we perform a displacement on one mode, $\varrho \rightarrow  D(\alpha) \varrho D^\dag (\alpha) \equiv \varrho_{\alpha}$, the effect on the characteristic function is easy to evaluate.
Using the relation $D (\alpha) D(\lambda)D^\dag(\alpha) = D(\lambda) \exp(-2i \mbox{Im}[\lambda \alpha^*]) $~\cite{Parisbook} we obtain
\begin{equation}
\chi [\varrho_{\alpha}] (\lambda_A, \lambda_B) =  \chi[\varrho] (\lambda_A, \lambda_B) \exp (-2 i \mbox{Im} [\lambda_B \alpha^* ] )
\end{equation}
Suppose we perform a masurement on mode $B$ in the number basis $\{ \Pi_n = \ket{n} \bra{n} \}$. The post-measurement state of mode $A$ is $ \varrho_{\alpha,n}^A = \frac{1}{p_n} \mbox{Tr}_B[\varrho_{\alpha} \Pi_{n}] $ where $p_n = \mbox{Tr} [\varrho_{\alpha} \Pi_{n}] $. By use of the trace formula~\cite{Parisbook}
\begin{equation*}
\mbox{Tr}[O_1 O_2] = \frac{1}{\pi} \int_{\mathbb{C}^m} d^{2m }\lambda \ \chi[O_1] (\lambda) \chi[O_2](-\lambda)
\end{equation*}
we obtain the characteristic function
\begin{equation}
\chi[\varrho_{\alpha_n}^A] (\lambda_A) = \frac{1}{\pi p_n }   \int_{\mathbb{C}} d^2 \lambda_B \ \chi[\Pi_{n}] (\lambda_B) \  \chi[\varrho_{\alpha}](\lambda_A, -\lambda_B)
\end{equation}
Since $ \chi[\Pi_{n}] (\lambda_B) = e^{-\frac{1}{2}|\lambda_B|^2} L_n (|\lambda_B|^2)$, where $L_n$ is the Laguerre polynomial $L_n (|\lambda_B|^2) = \sum_{i=0}^n \binom{n}{n-i}\frac{|\lambda_B|^{2i}}{i!}$, we have explicitly
\begin{eqnarray}
 \chi[\varrho_{\alpha_n}^A] (\lambda_A)   =  \frac{1}{\pi p_n }  \int_{\mathbb{C}} d^2 \lambda_B \  L_n (|\lambda_B|^2)  \exp\left( -  a |\lambda_A|^2 \right. \nonumber  \\  \nonumber
\left.   - (b + 1/2 ) |\lambda_B|^2 - 2 c \mbox{Re}[\lambda_A \lambda_B] - 2 i \mbox{Im} [\lambda_B \alpha^* ] \right)  \\ \label{Eq.: post-meas-state}
\
\end{eqnarray}
In order to see that this expression depends on $|\alpha|$ only we can implement the change
$\alpha \rightarrow \alpha' \equiv e^{ i \theta} \alpha $ and we have
\begin{eqnarray*}
 \chi[\varrho_{\alpha_n'}^A] (\lambda_A)  = \frac{1}{\pi  p_n } \int_{\mathbb{C}} d^2 \lambda_B \quad  L_n (|\lambda_B|^2) \exp\left( - a|\lambda_A|^2 \right.  \\ \nonumber
\left.  - (b + 1/2) |\lambda_B|^2 - 2 c \mbox{Re}[\lambda_A \lambda_B] - 2 i \mbox{Im} [\lambda_B \alpha^* e^{-i \theta} ] \right)
\end{eqnarray*}
By changing variable $\lambda_B \to e^{-i \theta} \lambda_B$ we see that
\be
  \chi[\varrho_{\alpha_n'}^A] (\lambda_A)  =
\chi[\varrho_{\alpha_n}^A] (\lambda_A e^{-i \theta})
\ee
Therefore, we have $ \chi[\varrho_{\alpha_n'}^A] \neq \chi[\varrho_{\alpha_n}^A] $, hence $ \varrho_{\alpha_n'}^A \neq \varrho_{\alpha_n}^A $. However, $\varrho_{\alpha_n'}^A$ and $\varrho_{\alpha_n}^A$ have the same spectrum. Indeed $ \chi[\varrho_{\alpha_n}^A(\lambda_A)] $ and
$ \chi[\varrho_{\alpha_n}^A(\lambda_A e^{i \theta})] $ are related by a simple quadrature rotation
\begin{equation}
a_1 \to a_1 e^{i \theta}  \quad a_1^\dag \to a_1^\dag e^{-i \theta}
\end{equation}
which means that
\begin{equation}
\varrho_{\alpha_n'}^A = U \varrho_{\alpha_n}^A U^\dag    \label{rotation}
\end{equation}
where $U$ is the free evolution of mode $A$, $U = e^{i \theta a_1^\dag a_1}$. Since $\varrho_{\alpha_n'}^A$ and $\varrho_{\alpha_n}^A$ are related by a unitary, they have the same spectrum.
Therefore, the spectrum (hence, the entropy) of the reduced post-measurement state $\varrho_{\alpha}^A$ does not depend on the phase of $\alpha$ but just on $|\alpha|$.  If follows that the non-Gaussian quantum discord of $\varrho_{\alpha}$ does not depend on the phase of $\alpha$, \emph{QED}. \\
As for the non-Gaussian geometric discord, it is obtained as
\begin{eqnarray*}
\mbox{Tr}[(\varrho)^2]-\sum_{n} \mbox{Tr}[ \varrho_{\alpha_n}^A \varrho_{\alpha_n}^A \otimes \Pi_{n}^B ]  =
\\ \nonumber
=\mbox{Tr}[(\varrho)^2]-\sum_{n} \mbox{Tr}[ \varrho_{\alpha_n}^A \varrho_{\alpha_n}^A ]
\end{eqnarray*}
By the same arguments before, leading to eq. (\ref{rotation}), we immediately see that the second trace
does not depend on the phase of $\alpha$, hence the geometric discord does neither.

\section{Understanding the behavior for growing $\alpha $}   \label{Sec : growing alpha}

\begin{figure}[bt]
\centering
\subfigure[Re]{\includegraphics[width=0.3\textwidth]{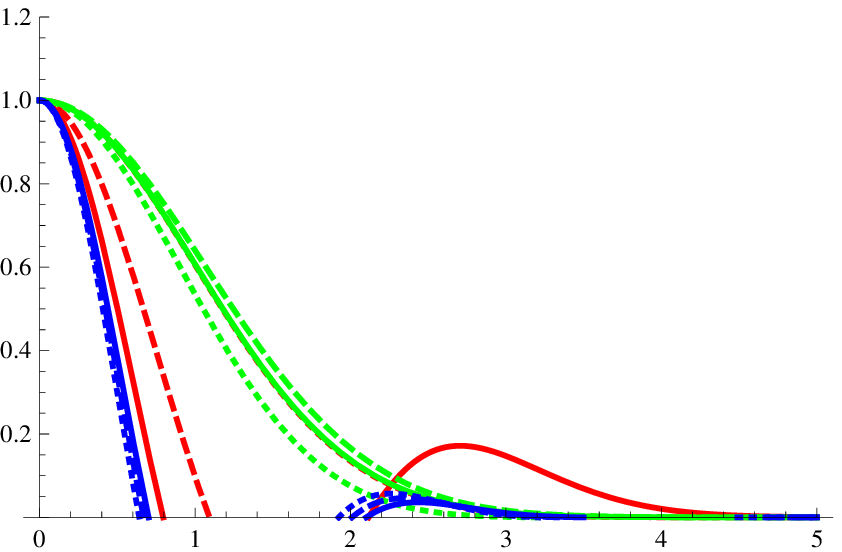}}
\subfigure[Im]{\includegraphics[width=0.3\textwidth]{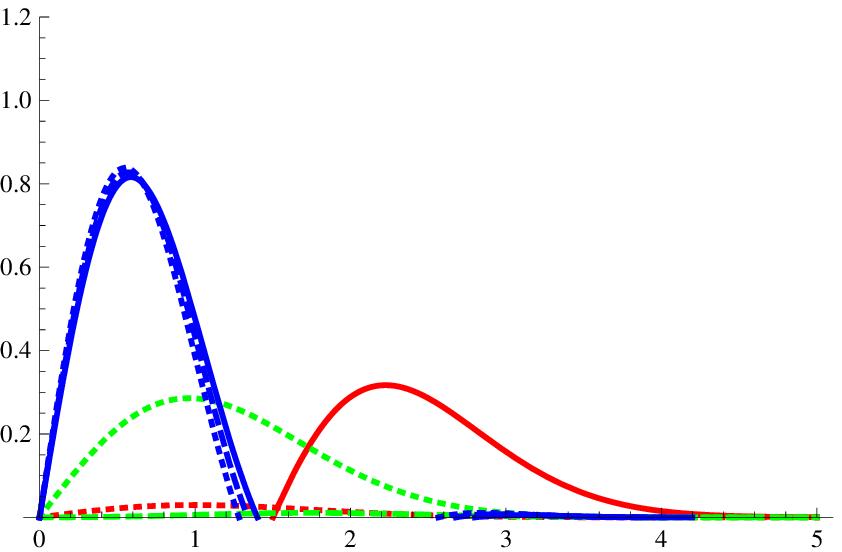}}
\caption{$\chi[\varrho_{\alpha_n^A}](\lambda)$ for $\alpha=0.1$ (red), $\alpha=1$ (green), $\alpha=5$ (blue) and $n=0$ (solid), $n=1$ (dashed), $n=2$ (dotted). Here, we have $A=( N_T +1/2 ) \cosh \lambda$, $C=( N_T +1/2 )\sinh \lambda$ with $\lambda=0.5$, $N_T = 0.5$.}
\label{Fig. : convergence}
\end{figure}

Let us now consider in detail the bahaviour for growing $\alpha $. We will show that the non-Gaussian discord in the displaced number basis tends to the Gaussian discord as the displacement tends to infinity, $D^{\mathcal{NG}} \to D^{\mathcal{G}}$ as $\alpha \to \infty$. \\
First, we will show that
\be
 \varrho_{\alpha_n}^A  \to \varrho_{\alpha_0}^A  \quad \mbox{as} \quad \alpha \to \infty  \label{Eq: limit2}
\ee
This is best shown in the characteristic function formalism.
The post-measurement state of mode $A$ has the characteristic function (\ref{Eq.: post-meas-state}).
Since the phase of $\alpha$ is irrelevant for the discord, we will assume $\alpha \in \mathbb{R}$ in the following.
The post-measurement state characteristic function, Eq. (\ref{Eq.: post-meas-state}), is the Gaussian integral of a polynomial. By using a well-known trick of Gaussian integrals, we can rewrite
\begin{eqnarray*}
& \ & \chi[\varrho_{\alpha_n}^A] (\lambda_A)  = \frac{1}{\pi p_n } e^{-a |\lambda_A|^2} \int_{\mathbb{C}} d^2 \lambda_B \quad  L_n (d/d\gamma) \times \\ \nonumber
& \ & \exp \left(- \gamma |\lambda_B|^2 + 2 c \mbox{Re} \lambda_A \mbox{Re} \lambda_B - (2 c \mbox{Im}[\lambda_A]  + 2 i \alpha) \mbox{Im} [\lambda_B]  \right)
\end{eqnarray*}
where $\gamma = b+1/2$ and the formal expression $ L_n(d/d \gamma) $  means $ \sum_{i=0}^n \binom{n}{n-i}\frac{1}{i!} \frac{d^n}{d\gamma^n}$. This expression can now be moved outside the integral, so that
we are now left with a purely Gaussian integral of the form
\begin{equation*}
\int_{\mathbb{R}} d^2 \mathbf{\Lambda}_B  \ e^{-\frac{1}{2} \mathbf{\Lambda}_B^T \mathcal{M} \mathbf{\Lambda}_B +  \mathbf{\Lambda}_B^T \mathcal{B} }
\end{equation*}
where $ \mathcal{M} = \mbox{diag}\{4 \gamma,4 \gamma\}$, $ \mathcal{B} = (2c\mbox{Re} \lambda_A, -2c\mbox{Im} \lambda_A + 2 i \alpha)$, $\mathbf{\Lambda_B} = (\mbox{Re} \lambda_B, \mbox{Im} \lambda_B) $.
The integral gives $\frac{2 \pi}{\sqrt{\mbox{det}\mathcal{M}}} e^{\frac{1}{2} \mathcal{B}^T \mathcal{M}^{-1}\mathcal{B}} $
so that we finally get
\begin{eqnarray}
\label{charfunc}
\chi[\varrho_{\alpha_n}^A] (\lambda_A)  = \frac{1}{p_n } e^{-\frac{1}{2} a |\lambda_A|^2 } L_n (-d/d\gamma) \times \\ \nonumber \frac{1}{\gamma} \exp\left( \frac{ c^2  |\lambda_A|^2 -  \alpha^2 - 2 i \alpha c \mbox{Im} \lambda_A }{2 \gamma} \right)
\end{eqnarray}
Let us define $ x =  c^2  |\lambda_A|^2 - \alpha^2 - 2 i \alpha c \mbox{Im} \lambda_A $. Then we have
\begin{eqnarray*}
L_n (-d/d\gamma) \frac{1}{\gamma} \exp\left( \frac{ c^2  |\lambda_A|^2 -  \alpha^2 - 2 i \alpha c \mbox{Im} \lambda_A }{2 \gamma} \right) & = & \\ \nonumber
= L_n (-d/d\gamma) \frac{1}{\gamma} e^{x/2\gamma} = F_n (\gamma, x) e^{x/2\gamma} & \ &
\end{eqnarray*}
where $F_n (\gamma, x) = \sum_k f_k (\gamma) x^k $ is necessarily a polynomial of degree $n$ in
with $\gamma$-dependent coefficients $f_k (\gamma)$. Therefore,
\begin{eqnarray}
\chi[\varrho_{\alpha_n}^A] (\lambda_A)  =  \frac{1}{p_n } e^{-(a - c^2(b+1/2)^{-1}) |\lambda_A|^2 } \times \\ \nonumber
e^{- i c(b+1/2)^{-1} \alpha Im \lambda_A} e^{-\alpha^2/2\gamma}
F_n (\gamma, x)
\end{eqnarray}
The norm is
\begin{eqnarray*}
 p_n = e^{- (a - c^2(b+1/2)^{-1}) |\lambda_A|^2 } e^{- i c(b+1/2)^{-1} \alpha Im \lambda_A} \times \\ \nonumber
e^{-\alpha^2/2\gamma} F_n (\gamma, x) \Big|_{\lambda_A=0} = e^{-\alpha^2/2\gamma} F_n (-\alpha^2)
\end{eqnarray*}
so that
\begin{eqnarray}
\chi[\varrho_{\alpha_n}^A] (\lambda_A)  = e^{- (a - c^2(b+1/2)^{-1}) |\lambda_A|^2 }  \times \\ \nonumber
e^{- i c(b+1/2)^{-1} \alpha Im \lambda_A} \frac{F_n (\gamma, x)}{F_n (\gamma, -\alpha^2) }
\end{eqnarray}
This function is exponentially decaying as $ e^{- s |\lambda_A|^2} $ where $s = a - c^2/(b+1/2) $, hence
it is vanishing for $ |\lambda_A|^2 \gg 1/s $.  Therefore, we can consider values of $|\lambda_A|^2$ in the region $\lambda_A^2 \lesssim 1/s $. In this region, we we have $\lim_{\alpha \to \infty} x =  -\alpha^2$
because $\alpha \gg \lambda_A$ and thus
\begin{equation*}
\lim_{\alpha \to \infty}  \frac{F_n (\gamma, x)}{F_n (\gamma, -\alpha^2) } = \frac{f_n (\gamma) \alpha^2n }{f_n (\gamma) \alpha^2n } = 1
\end{equation*}
In conclusion, as $\alpha \to \infty $ we have
\begin{eqnarray}
& \ & \chi[\varrho_{\alpha_n}^A] (\lambda_A)  \to \chi[\varrho_{\alpha_0}^A]
\end{eqnarray}
which implies the desired result (\ref{Eq: limit2}), \emph{QED}. \par

This result means that the conditional state of $A$ is independent of $n$ and equal to the $n=0$ result.
In fig. \ref{Fig. : convergence} we show $\chi[\varrho_{\alpha_0}^A], \chi[\varrho_{\alpha_1}^A],\chi[\varrho_{\alpha_2}^A] $  for  growing values of $\alpha$. The three curves converge already for  $\alpha \sim 5$.
As a consequence of $n$-independence, we have
\begin{equation}
S^{\Pi,\mathcal{NG}} (A|B) = \sum_n   p_n S(\varrho_{\alpha_n}^A) \to S(\varrho_{\alpha_0}^A) 
\end{equation}
But $\varrho_{\alpha_0}^A $ is just the post-measurement state corresponding to POVM element $D(\alpha) \ket{0} \bra{0} D^\dag (\alpha)=\ket{\alpha} \bra{\alpha}$, i.e, a Gaussian state with covariance matrix   $\sigma_P = A -C(B+\mathbb{I}/2)^{-1} C^T $ (Schur complement), and mean $  \mu_P = X (B+\mathbb{I}/2)^{-1} C^T  $,
where $ X = (\alpha, 0)$. On the other hand, from the discussion in sec. \ref{Sec : correlations} we know that the optimal Gaussian POVM is a heterodyne measurement $\{ \Pi_{\beta} = D(\beta) \ket{0} \bra{0} D^\dag(\beta) = \ket{\beta} \bra{\beta} \} $. In this case, as already explained in sec.~\ref{Sec : correlations}, the entropy of the post measurement state $\varrho_{\beta}^A $ is independent of the measurement result $\beta$, hence the conditional entropy coincides with the entropy of of the $\beta=\alpha$ result. Therefore, we also have
$ S^{\mathcal{G}} (A | B) = S(\varrho_{\alpha_0})$.
Therefore, we conclude that the non Gaussian discord $D^{\mathcal{NG}}(A:B)$ in the displaced number basis tends to the Gaussian discord $D^{\mathcal{G}}(A:B)$ as $\alpha \to \infty$, \emph{QED}. \\
To be rigorous, we did not prove that the $D^{\mathcal{NG}}(A:B)$ is lower bounded by  $D^{\mathcal{G}}(A:B)$, and we cannot rule out the possibility that $D^{\mathcal{NG}}(A:B) <  D^{\mathcal{G}}(A:B)$  for intermediate values of $\alpha$. However, our numerical data do not support this possibility since we never observe $D^{\mathcal{NG}}(A:B) <  D^{\mathcal{G}}(A:B)$ and we expect that $D^{\mathcal{NG}}(A:B) \to  D^{\mathcal{G}}(A:B)$ from above as $\alpha \to \infty$.


\begin{thebibliography}{30}
\bibitem{Horodecki} R. Horodecki, P. Horodecki, M. Horodecki, 
and K. Horodecki, Rev. Mod. Phys. \textbf{81}, 865  (2009).
\bibitem{Ollivier} H. Ollivier and W. H. Zurek, Phys. Rev. Lett. \textbf{88}, 017901 (2001).
\bibitem{Vedral} L. Henderson and V. Vedral, J. Phys. A \textbf{34}, 6899 (2001).
\bibitem{DiscordRev} K. Modi, A. Brodutch, H. Cable, T. Paterek and V. Vedral, arXiv:1112.6238 (2011).
\bibitem{Gu} M. Gu, H. M. Chrzanowski, S. M. Assad, T. Symul, K. Modi, T. C. Ralph, V. Vedral, P. K. Lam,
arXiv:1203.0011 (2012).
\bibitem{Datta2} V. Madhok, A. Datta, arXiv:1204.6042 (2012).
\bibitem{Datta} A. Datta, A. Shaji, and C. M. Caves, Phys. Rev. Lett. \textbf{100}, 050502 (2008);
A. Datta and A. Shaji, Int. J. Quant. Inf. \textbf{9}, 1787 (2011);
A. Brodutch and D. R. Terno, Phys. Rev. A \textbf{83}, 010301 (2011);
A. Al-Qasimi and D. F. V. James, Phys. Rev. A \textbf{83}, 032101 (2011);
G. Passante, O. Moussa, D. A. Trottier, R. Laflamme, Phys. Rev. A. \textbf{84}, 044302 (2011)  (2011).
\bibitem{RSP} B. Daki\'c \textit{et al.}, arXiv:1203.1629 (2012).
\bibitem{Fer12} A. Ferraro, M. G. A. Paris, arXiv:1203.2661 (2012).
\bibitem{Dakic} B. Daki\'c, C. Brukner, and V. Vedral, Phys. Rev. Lett. \textbf{105}, 190502 (2010).
\bibitem{AnalyticDiscord} S. Luo, Phys. Rev. A \textbf{77}, 042303 (2008); M. Ali, A. R. P. Rau, and G. Alber, Phys. Rev. A \textbf{81}, 042105 (2010).
\bibitem{GiordaGaussDiscord} P. Giorda and M. G. A. Paris, Phys. Rev. Lett. \textbf{105}, 020503 (2010).
\bibitem{AdessoDatta} G. Adesso and A. Datta, Phys. Rev. Lett. \textbf{105}, 030501 (2010).
\bibitem{GauuPOVM} G. Giedke and J. I. Cirac, Phys. Rev. A \textbf{66}, 032316 (2002); J. Fiur\`a\v{s}ek and L. Mi\v{s}ta Jr., Phys. Rev. A (\textbf{75}, 060302(R) 2007).
\bibitem{Vasile} R. Vasile, P. Giorda, S. Olivares, M. G. A. Paris, and S. Maniscalco, Phys. Rev. A \textbf{82}, 012313 (2010); G. L. Giorgi, F. Galve, R. Zambrini, Int. J. Quant. Inf. \textbf{9}, 1825 (2011); L. A. Correa, A. A. Valido, D. Alonso. arXiv:1111.0806v2 (2011).
\bibitem{Zambrini} G. L. Giorgi, F. Galve, G. Manzano, P. Colet, R. Zambrini, Phys. Rev. A \textbf{85}, 052101 (2012).
\bibitem{NonGausDisc} R. Tatham, L. Mi\v{s}ta Jr., G. Adesso, N. Korolkova, Phys. Rev. A \textbf{85}, 022326 (2012).
\bibitem{Koashi} M. Koashi and A. Winter, Phys. Rev. A {\bf69}, 022309 (2004).
\bibitem{GaussianGeom} G. Adesso and D. Girolami, Int. J. Quant. Inf. \textbf{9} (2011).
\bibitem{EOF} G. Giedke, M. M. Wolf, O. Kr\"uger, R. F. Werner, and J. I. Cirac, Phys. Rev. Lett. \textbf{91}, 107901 (2003); J. Solomon Ivan, R. Simon,  arXiv:0808.1658; P. Marian, T. A. Marian, Phys. Rev. Lett. \textbf{101}, 220403 (2008).
\bibitem{DO} M. G. A. Paris, Phys. Lett. A \textbf{217}, 78 (1996).
\bibitem{DisplacedNumber} 
F. A. M. de Oliveira, M. S. Kim, P. L. Knight, abd V. Bu\v{z}ek, 
Phys. Rev. A \textbf{41}, 2645 (1990).
\bibitem{SNS1}  
H. P. Yuen, J. Opt. Soc. Am B {\bf 3}, P86 (1986).
\bibitem{SNS2}  
M. S. Kim, F. A. M. de Oliveira, and P. L. Knight, Phys. Rev. A \textbf{40}, 
2494 (1989).
\bibitem{SNS3} 
R. T. Hammond, Phys. Rev. A {\bf 41}, 1718 (1990).
\bibitem{SNS4}  
C. F. Lo , Phys. Rev. A \textbf{43}, 404 (1991); 
M. M. Nieto, Phys. Lett. A \textbf{229}, 135 (1997).
\bibitem{STSteo} C. T. Lee, Phys. Rev. A \textbf{42}, 4193 (1990).
\bibitem{STSexp1} 
A. Furusawa, J. L. Sorensen, S. L. Braunstein, C. A.
Fuchs, H. J. Kimble, and E. S. Polzik, Science {\bf 282}, 706
(1998); N. Lee, H. Benichi, Y. Takeno, S. Takeda, J. Webb, E.
Huntington, A. Furusawa, Science, {\bf 332},330 (2011).
\bibitem{STSexp2} 
V. D'Auria, S. Fornaro, A. Porzio, S. Solimeno, 
S. Olivares, and M. G. A. Paris, Phys. Rev. Lett. \textbf{102}, 020502 (2009).
\bibitem{MTS} A. Agliati, M. Bondani, A. Andreoni, G. De Cillis, M. G.
A. Paris, J. Opt. B {\bf 7}, 652 (2005).
\bibitem{SingleModeSqueezing}
L. Albano, D.F.Mundarain  and J. Stephany, J. Opt. B \textbf{4}, 352 (2002).
\bibitem{SingleModeSqueezingKnight}
M. S. Kim, F. A. M. De Oliveira and P. L. Knight, Phys. Rev. A \textbf{40}, 2494 (1989).
\bibitem{Parisbook} A. Ferraro, S. Olivares, M. G. A. Paris, \textit{Gaussian states in continuous variable quantum information}, Bibliopolis, Napoli (2005), arXiv:quant-ph/0503237.
\bibitem{Luo} S. Luo and S. Fu, Phys. Rev. A \textbf{82}, 034302 (2010).
\bibitem{Tufarelli} T. Tufarelli, D. Girolami, 
R. Vasile, S. Bose, G. Adesso,  arXiv:1205.0251.
\bibitem{Piani} M. Piani, arXiv:1206.0231 (2012).
\end{thebibliography}
\end{document}